\documentclass[a4paper,fleqn,usenatbib]{mnras}

\usepackage{newtxtext,newtxmath}

\usepackage[T1]{fontenc}
\usepackage{ae,aecompl}

\usepackage{graphicx}	
\usepackage{amsmath}	
\usepackage{amssymb}	

\usepackage{rotating}
\usepackage{threeparttable}

\usepackage{dcolumn}
\usepackage{longtable}
\usepackage{multirow}
\usepackage{bm}
\usepackage{textpos}

\title[Numerical Modelling of Tertiary Tides]{Numerical Modelling of Tertiary Tides}

\author[Y. Gao et al.]{
Yan Gao$^{1,2}$\thanks{E-mail: ygbcyy@ynao.ac.cn},
Alexandre C.M. Correia$^{3,4,5}$,
Peter P. Eggleton$^{6}$
and Zhanwen Han$^{1,2}$
\\
$^{1}$Yunnan Observatories, Chinese Academy of Sciences, Kunming 650011, China\\
$^{2}$Key Laboratory for the Structure and Evolution of Celestial Objects, Chinese Academy of Sciences, Kunming 650011, China\\
$^{3}$Department of Physics, University of Coimbra, 3004-516 Coimbra, Portugal\\
$^{4}$CIDMA, Department of Physics, University of Aveiro, 3810-193 Aveiro, Portugal\\
$^{5}$ASD, IMCCE, Paris Observatory, PSL University, 77 Av. Denfert-Rochereau, 75014 Paris, France\\
$^{6}$Lawrence Livermore National Laboratory, 7000 East Ave, Livermore, CA 94551, USA
}

\date{Accepted for publication in MNRAS}

\pubyear{2018}

\begin{document}
\label{firstpage}
\pagerange{\pageref{firstpage}--\pageref{lastpage}}
\maketitle

\begin{abstract}

Stellar systems consisting of multiple stars tend to undergo tidal interactions when the separations between the stars are short. While tidal phenomena have been extensively studied, a certain tidal effect exclusive to hierarchical triples (triples in which one component star has a much wider orbit than the others) has hardly received any attention, mainly due to its complexity and consequent resistance to being modelled. This tidal effect is the tidal perturbation of the tertiary by the inner binary, which in turn depletes orbital energy from the inner binary, causing the inner binary separation to shrink. In this paper, we develop a fully numerical simulation of these "tertiary tides" by modifying established tidal models. We also provide general insight as to how close a hierarchical triple needs to be in order for such an effect to take place, and demonstrate that our simulations can effectively retrieve the orbital evolution for such systems. We conclude that tertiary tides are a significant factor in the evolution of close hierarchical triples, and strongly influence at least $\sim1\%$ of all multiple star systems.
\end{abstract}

\begin{keywords}
celestial mechanics, (stars:) binaries (including multiple): close, stars: evolution
\end{keywords}



\section{Introduction}

Stars in close multiple systems are subject to tidal forces, which play a pivotal role in shaping their futures. The intrinsic mechanism behind these tidal forces is that, for every celestial body within a multiple system, the motion of the other bodies subjects it to a changing gravitational field, inducing internal motion within it, which in turn affects the gravitational field emanating from it, thereby influencing the rest of the system as a whole. In conjunction with dissipative processes (see \citealt{2014ARA&A..52..171O} for a review of such processes), tidal forces facilitate, among many other effects, the migration of angular momentum from one part of the system to another. Due to the importance of the various roles of these forces, previous studies have conducted extensive investigations about their nature.

Despite the relative simplicity of the concept, clarity has yet to be achieved as to exactly how tidal forces ought to be modelled. Some researchers \citep[e.g.][]{1981A&A....99..126H,1998ApJ...499..853E,1998MNRAS.300..292K, 2016CeMDA.126..189C} favour a treatment based on equilibrium tides (usually referred to as the "equilibrium tide model"), while others \citep[e.g.][]{1977ApJ...213..183P,1995ApJ...450..722M,1995ApJ...450..732M, 1996ApJ...466..946K, 2014A&A...571A..50C, 2017CeMDA.128...19R} advocate a treatment that approximates the celestial body receiving the tidal force as an oscillator with many different oscillation modes, each one absorbing energy in its own way (known as the "dynamical tide model"). It has been pointed out that the two models may be complementary \citep[e.g.][]{1998ApJ...499..853E}, with each model being optimized for a special set of cases, but even so, it is still unclear where the line should be drawn when dealing with specific systems.

Yet however great the controversy may be when it comes to modelling tidal processes, there is a general consensus regarding the macroscopic effects of tidal forces; that they tend to synchronise the rotations and orbits of all the bodies involved, circularise orbits by causing a decay in their ellipticities, and convert certain portions of the kinetic and potential energies of the bodies involved into heat, which can then be radiated away. For instance, for a 2-body system in a Keplerian orbit under tidal effects, given time, the system must ultimately evolve into a circular orbit, with the orbital angular velocity being equal to the respective rotational angular velocities of each body, regardless of how eccentric their initial orbit may be or how much their initial angular velocities may differ. 

\begin{figure}
\includegraphics[scale=0.52, angle=270]{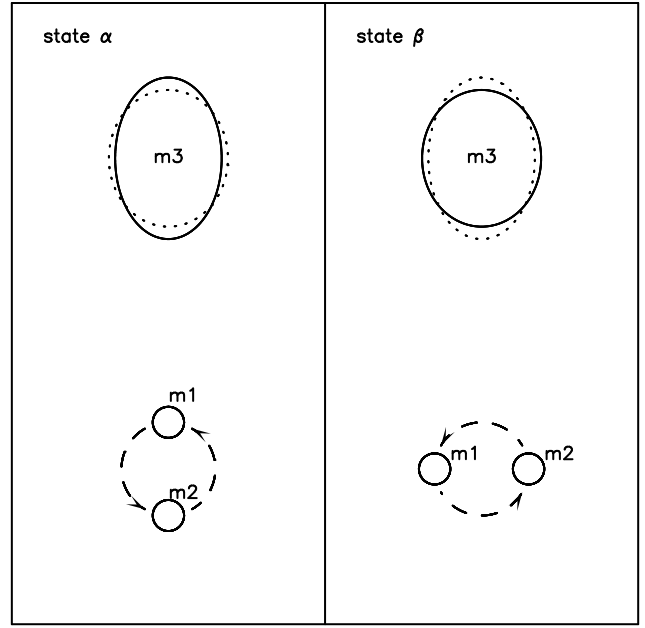}
\caption{Illustration of how the inner binary affects the third body when tertiary tides become significant (see also the animated figure attached to this paper). The state when all three bodies are aligned, as depicted in the left panel, is defined as state $\alpha$, and the state in which the three bodies are at the vertices of an isosceles triangle, as depicted in the right panel, is defined as state $\beta$. The solid lines display the shape of the tertiary at equilibrium tidal distortion, while the dotted lines represent the same shape in the other state for comparison. The tidal distortion of the tertiary is greatly exaggerated. \label{Fig1}}
\end{figure}

Of all the tidal effects to which close multiple systems are exposed, only three remain relevant for the orbital evolution of a hierarchical triple system (consisting of an inner binary of masses $m_{\rm 1}$ and $m_{\rm 2}$, as well as an outer tertiary of mass $m_{\rm 3}$). The first is the tidal locking between $m_{\rm 1}$ and $m_{\rm 2}$, which is no different from 2-body tidal effects in general, and has historically been the subject of intense study \citep[e.g.][]{1973ApJ...180..307C,1981A&A....99..126H}. The second is the tidal locking of $m_{\rm 3}$ to the inner binary, which will eventually synchronize the rotation and the orbit of $m_{\rm 3}$ \citep[e.g.][]{2016CeMDA.126..189C}. The third and final effect is the dumping of energy from $m_{\rm 1}$ and $m_{\rm 2}$ to $m_{\rm 3}$, which $m_{\rm 1}$ and $m_{\rm 2}$ achieve by tidally distorting $m_{\rm 3}$ as illustrated in the cartoon depiction in Fig. \ref{Fig1} (see also Animated Figure 1). 

As portrayed in Fig. \ref{Fig1}, $m_{\rm 3}$ receives the greatest amount of tidal force from $m_{\rm 1}$ and $m_{\rm 2}$ when all 3 bodies are aligned (left panel), and receives the least when they are orthogonal (right panel). This change in received tidal force translates into a change in the degree of tidal distortion (elongation in the the direction of $m_{\rm 1}$ and $m_{\rm 2}$) that $m_{\rm 3}$ undergoes. Consequently, if the internal tidal friction in $m_{\rm 3}$ is strong enough to (at least partly) brake the resultant internal motion, this leads to (at least part of) the energy carried in the tidal distortion difference being converted to heat. At whatever rate this process generates heat, it must essentially be fuelled by the orbital energy within the orbit of $m_{\rm 1}$ and $m_{\rm 2}$, which is the driving motion behind the tidal distortion of $m_{\rm 3}$. Therefore, this effect should serve also to drive the inner binary separation to be smaller. Here, it should be noted that these tidal effects will not end in tidal locking, as is often the case with 2-body tidal effects, since no rotation of $m_{\rm 3}$ can decrease the difference in self-gravitational potential energy in the transition between the left and right panels of Fig. \ref{Fig1}. 


Of the three effects mentioned above, the first two have already been extensively investigated, as is evident from the literature. Very little attention, however, has been paid to the third. Admittedly, this is not entirely without good reason; in a vast majority of cases, $m_{\rm 3}$ is much smaller than its Roche Lobe, and the tidal distortion it undergoes is consequently insignificant. However, whenever the condition comes to pass that $m_{\rm 3}$ is more or less the same size as its Roche Lobe, this third effect becomes interesting for one simple reason: as mentioned above, this third effect can never be mitigated by tidal locking, and therefore can theoretically form an endless drain of the orbital energy of $m_{\rm 1}$ and $m_{\rm 2}$. Furthermore, we shall show that, unlike any other merger-contributing mechanism investigated so far, {\it the greater the inner binary separation, the greater this energy drain per unit time will be}. In other words, this effect is rare in that it preferentially allows large binary separations to decrease. So far, triple systems, in which $m_{\rm 3}$ is close enough for this third effect to have been prominent in the past, have occasionally been observed \citep[e.g.][]{2011Sci...332..216D}. Speculation has also arisen that the inner binaries have been driven closer together due to 3-body tidal effects, which are not inconsistent with observed properties of these triples \citep[e.g.][]{2013MNRAS.429.2425F}. However, there is not, to the knowledge of the authors, as of yet any work that provides a simulation which can recover the exact details of this third effect, and therefore the way in which the orbits of triple systems under its influence evolve is not well understood. We seek to remedy this.

For the rest of this paper, we shall refer to this third effect by the names ``tertiary tides" or ``TTs" for short.
In what follows in this paper, we describe our model and its numerical implementation in \S 2, and present the results of our calculations for some specific systems in \S 3. Finally, our conclusions regarding the influence of tertiary tides in general, as well as the limitations of our work, are provided in \S 4, along with an extrapolation of what work could be done in the future.



\section{Treatment of Tides and Tidal Lags}

To reliably simulate a close triple system undergoing TTs, we adopt a 2-stage simulation based on 8th-order Runge Kutta methods (hereafter RK8), by modelling the orbital motion of three bodies in a hierarchical triple configuration. We treat the bodies constituting the inner binary $m_{\rm 1}$ and $m_{\rm 2}$ as point masses, whereas the third body $m_{\rm 3}$ is modelled as a body with a gravitational field varying with time, in order to account for its tidal distortion. In the first stage, we calculate the amount of energy extracted from the inner orbit per unit time, in which TTs are taken into account by means of a modified version of classical 2-body tidal models \citep[e.g.][]{1973ApJ...180..307C,1981A&A....99..126H,1977A&A....57..383Z,2005ASPC..333....4Z}. For the second stage, we adopt a viscoelastic tidal model \citep[e.g.][]{2014A&A...571A..50C}, calibrating an unknown parameter $\tau$ in this model by varying the parameter until the energy extraction rates of the two models match. This provides detailed positions and velocities of all 3 bodies, as well as the rotation and deformation of $m_{\rm 3}$, as a function of time. From these positions and velocities, orbital parameters (such as semimajor axes and periods) of both the inner and outer orbits can be retrieved. The details of each of these two stages are presented below.

\subsection{Stage 1 Simulations}
\label{s1ss1}

We consider the special case of a hierarchical triple, consisting of a double point mass inner binary ($m_{\rm 1}$, $m_{\rm 2}$) in a circular orbit, and a coplanar third body ($m_{\rm 3}$), also in a circular orbit around the center of mass (COM) of the inner binary (Jacobi coordinates). For simplicity, we assume that $m_{\rm 1}=m_{\rm 2}$, and that $m_{\rm 3}$ is tidally locked to the inner binary's COM - in such a system, we do not consider rotational effects. Had it been the case that internal dissipation within $m_{\rm 3}$ were very efficient, the rate at which orbital energy is dumped from the inner binary to $m_{\rm 3}$ can be shown to be
\begin{equation}
{\frac{{\rm d}E}{{\rm d}t}}{\sim}{\frac{135}{4}}\frac{Gm^{2}R_{\rm 3}^{5}a_{\rm 1}^{2}}{a_{\rm 2}^{8}}{\frac{4}{P_{\rm in}}},
\label{delta_p_homo}
\end{equation}
via a set of trivial calculations (see appendix \ref{app_other} for details). Here, $m=m_{\rm 1}=m_{\rm 2}$, $R_{\rm 3}$ is the radius of $m_{\rm 3}$, $a_{\rm 1}$ and $a_{\rm 2}$ are the semi-major axes of the inner and outer orbits respectively, and $P_{\rm in}$ is the inner orbital period. However, since dissipation efficiency might be very low for this process (and we indeed find it to be so in our work), we need a much more detailed set of simulations, detailed below. 

At each moment, $m_{\rm 3}$ has a proclivity to assume the distortion corresponding to the equilibrium tide of that particular moment. This equilibrium distortion at each moment leads to a gravitational field that can be approximately expressed by (see appendix \ref{delayed_to_visco})
\begin{equation}
V\left(r,{\psi}\right)=-\frac{Gm_{\rm 3}}{r} \left[1+k_{\rm 2} \, \zeta(\phi)\left(\frac{R_{\rm 3}}{r}\right)^{2}P_{\rm 2}(\cos{\psi}) \right],
\label{phi}
\end{equation}
\noindent where $k_{\rm 2}$ is the Love number for $m_{\rm 3}$ (for polytropic stars with $n=1.5$, we use $k_{\rm 2}=0.2$ as an approximation, following \citealt{2017MNRAS.472.4965Y}), $r$ is the distance measured from the center of $m_{\rm 3}$, the angle $\psi$ is defined to be zero in the direction of the tertiary bulge maximum, and ${\zeta}$ is the tidal distortion parameter, 
\begin{equation}
\zeta (\phi) = \left[\frac{P_{\rm 2}(\cos{\psi_1})}{(r_{\rm 1}/a_{\rm 2})^3}+\frac{P_{\rm 2}(\cos{\psi_2})}{(r_{\rm 2}/a_{\rm 2})^3}\right]
\frac{m}{m_{\rm 3}}\left(\frac{R_{\rm 3}}{a_{\rm 2}}\right)^3
, \label{zetaphi}
\end{equation}
\noindent where $r_{\rm 1}$ and $r_{\rm 2}$ are the distances from $m_{\rm 1}$ to $m_{\rm 3}$ and $m_{\rm 2}$ to $m_{\rm 3}$, respectively, ${\phi}$ is the angle between $\overrightarrow{m_{\rm 1}m_{\rm 2}}$ and $\overrightarrow{m_{\rm 3}C}$ ($C$ being the COM of the inner binary), and $\psi_1$ and $\psi_2$ are the $\psi$ values for $m_{\rm 1}$ and $m_{\rm 2}$ respectively.
For circular orbits and $\alpha = a_1/a_2$, 
\begin{equation}
\begin{split}
\left(\frac{r_{\rm 1}}{a_{\rm 2}}\right)^3&=\left(1+\alpha{\cos\phi}+\frac{1}{4}\alpha^2\right)^{3/2} \ ,\\
\left(\frac{r_{\rm 2}}{a_{\rm 2}}\right)^3&=\left(1-\alpha{\cos\phi}+\frac{1}{4}\alpha^2\right)^{3/2} \ .
\end{split}
\end{equation}
\noindent Since internal dissipation is not instantaneous, the tertiary never achieves this equilibrium. Instead, it assumes some tidal distortion equivalent to its equilibrium state a certain amount of time $t_{\rm lag}$ ago, where $t_{\rm lag}$ is usually termed the tidal lag time.

But how much time is $t_{\rm lag}$? To answer this question, we draw an analogy from the $t_{\rm lag}$ in binary tidal locking. In a binary with component stars $A$ and $B$ (totally different from and irrelevant to the triple system mentioned above), where star $A$ is an extended object which is not rotating, and star $B$ is a point mass orbiting star $A$ in a circular orbit at an angular velocity of $\omega$, binary tidal locking occurs as follows. The existence of star $B$ is supposed to distort star $A$ at every epoch in a way such that star $A$ is elongated in the direction of star $B$, forming two bulges on its surface, one pointing towards and the other away from star $B$, if equilibrium tides are assumed. However, since star $A$ undergoes internal friction due to viscous processes, the orientation of those bulges always lags behind the orientation corresponding to equilibrium tide, or in other words, star $A$ is constantly in a state of distortion corresponding to its tidal equilibrium a certain amount of time $t_{\rm lag}$ ago. Thus, the bulges are always aligned towards a position that star $B$ was the same amount of time $t_{\rm lag}$ ago, corresponding to an angle $\lambda$ away from $B$, thereby introducing a torque on the orbit of star $B$, decreasing its orbital angular momentum. Conversely, the rotational angular momentum of star $A$ must increase due to conservation of angular momentum throughout the system.

It has been shown that $t_{\rm lag}$ can be expressed as
\begin{equation}
t_{\rm lag}=\frac{P}{2{\pi}}{\lambda},
\label{tlag}
\end{equation}
\noindent where $P$ is the orbital period of the binary, and $\lambda$ is the tidal lag angle, which can be expressed as \citep{1977A&A....57..383Z,2005ASPC..333....4Z}
\begin{equation}
{\lambda}={\omega}t_{\rm dyn}^{2}\left(\frac{1}{t_{\rm diss}}\right)
\label{lag_angle}
\end{equation}
\noindent where $t_{\rm dyn}$ is the dynamical timescale of star $A$, $\omega$ is the orbital angular velocity described above, and $t_{\rm diss}$ is the typical dissipation timescale of star $A$. According to its definition, $t_{\rm dyn}$ is simply
\begin{equation}
t_{\rm dyn}=\sqrt{\frac{R^3}{GM}}.
\label{tdyn}
\end{equation}
\noindent The value of $t_{\rm diss}$, however, is a somewhat more complicated issue, and here we focus only on the aspects of its calculation immediately relevant to this paper. For a star with a convective envelope (which is the case for both low-mass main sequence stars and red giants), turbulent convection dominates the dissipation process for equilibrium tides \citep[e.g.][]{2005ASPC..333....4Z}, and $t_{\rm diss}$ is simply the convective timescale $t_{\rm conv}$ when the tidal period (the period of variation of the tidal forcefield, which is also the orbital period in a 2-body scenario) is longer than $t_{\rm conv}$:
\begin{equation}
t_{\rm diss}=t_{\rm conv}=\left(\frac{MR^2}{L}\right)^{1/3}.
\label{tconv}
\end{equation}
\noindent where $M$, $R$ and $L$ are the mass, radius and luminosity of star $A$, respectively. However, it is important to note that the above equation is only valid when the tidal forcefield changes very slowly, giving the perturbed body ample time to dissipate energy, as is the case when the tidal period is longer than $t_{\rm conv}$. When the tidal period is shorter than $t_{\rm conv}$, the perturbed body doesn't have sufficient time to dissipate this energy before the tidal forcefield reverts back to its former state, in which case a phenomenon called ``fast tides" starts to come into effect. When this happens, $t_{\rm diss}$ ought to be calculated via either
\begin{equation}
t_{\rm diss}=\left(\frac{t_{\rm conv}}{P}\right)t_{\rm conv},
\label{tdiss1}
\end{equation}
\noindent or
\begin{equation}
t_{\rm diss}=\left(\frac{t_{\rm conv}}{P}\right)^{2}t_{\rm conv},
\label{tdiss2}
\end{equation}
\noindent or perhaps
\begin{equation}
t_{\rm diss}=\left(\frac{t_{\rm conv}}{P}\right)^{5/3}t_{\rm conv},
\label{tdiss3}
\end{equation}
\noindent according to \citet{2005ASPC..333....4Z}, \citet{1977ApJ...211..934G}, and \citet{1997ApJ...486..403G}, respectively. It is not currently known which, if any at all, of these treatments approximates fast tides well \citep[e.g.][]{2014ARA&A..52..171O,2016A&A...592A..33M}, but recent results seem to favour the first prescription for stellar interiors \citep{2007ApJ...655.1166P,2009ApJ...704..930P}, and hence we will use this prescription for our following calculations.

Having found a way to calculate $t_{\rm lag}$ for binary tides, we return to our previous triple system with $m_{\rm 1}$, $m_{\rm 2}$, and $m_{\rm 3}$. The method above can be converted into a calculation for $t_{\rm lag}$ in TTs as shown below.

\begin{figure}
\includegraphics[scale=0.52, angle=270]{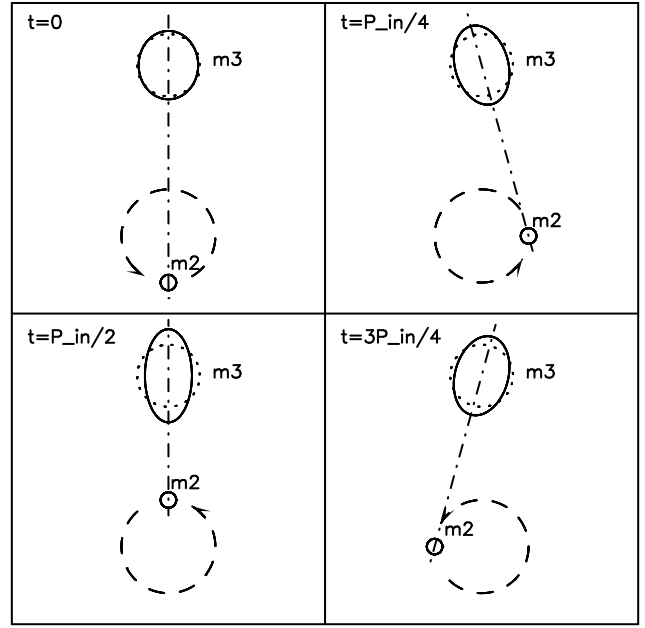}
\caption{Dissection of tertiary tides - how the tertiary is tidally distorted in reaction to the tidal forcing from one component of the inner binary alone. Here, the effects of the other component of the inner binary have been eliminated, but its companion is assumed to travel in the same orbit as before. The solid lines display the shape of the tertiary at equilibrium tidal distortion, while the dotted lines represent its shape when no tidal forces are applied. The dash-dotted lines indicate the direction of elongation of the equilibrium tide distortion, which is invariably in the direction of the perturbing body. The time given in the upper left corner of each panel is given in units of inner binary orbital period. It can be seen that the net result is an oscillatory rotational effect, but this rotation is largely cancelled out by the effects of the other inner binary component when full tertiary tides are considered. The tidal distortion of the tertiary is greatly exaggerated. \label{Fig2}}
\end{figure}

The tidal distortion of $m_{\rm 3}$ experienced during TTs depicted in Fig. \ref{Fig1} is a combination of separate tidal distortions by $m_{\rm 1}$ and $m_{\rm 2}$ (see Fig. \ref{Fig2} for an illustration of a single component). Since $m_{\rm 1}=m_{\rm 2}$, and considering the general symmetry of the inner binary, the $t_{\rm lag}$ of the distortion caused by $m_{\rm 1}$ must be equal to that caused by $m_{\rm 2}$. Hence, we only need to calculate the value of $t_{\rm lag}$ due to either $m_{\rm 1}$ or $m_{\rm 2}$ in order to retrieve the $t_{\rm lag}$ for the tidal distortion of $m_{\rm 3}$.

How, then, should this $t_{\rm lag}$ be calculated, and how long is it for the triple system in question? To answer this, we revert to the calculation of the tidal lag time in a binary system, as presented in Eqs. \ref{tlag} and \ref{lag_angle}. For our TTs, the tertiary lag time should also be calculable with these equations, albeit with minor modifications as to what physical quantities each of the variables correspond to in a triple system under TTs. The dynamical timescale $t_{\rm dyn}$ is indisputably the dynamical timescale of the tertiary, but for the other variables, namely $P$, $\omega$, and $t_{\rm diss}$, it may not be so obvious.

To find what value to substitute for $P$, one must discern exactly what $P$ is in Eq. \ref{tlag}. Considering that $\lambda$ is simply the tidal lag angle, and hence whatever remains is merely a conversion factor from lag angle to lag time, it becomes clear that $P$ is more related to how the perturbing body is moving relative to the perturbed body than it is to the intrinsic period of the acting tidal force. In other words, it would not matter at any particular moment if the perturbing body were travelling in a circular orbit, or in a straight line tangential to that circular orbit, and had happened to be at the point of intersection at that particular moment - both scenarios would result in the same $t_{\rm lag}$, had the lag angle been the same. In fact, Eqs. \ref{tlag} and \ref{lag_angle} could be more accurately expressed as
\begin{equation}
t_{\rm lag}=\frac{d}{v}\left({\omega}t_{\rm dyn}^{2}\frac{1}{t_{\rm diss}}\right).
\label{tlag_alt}
\end{equation}
\noindent where $d$ is the distance between the perturbed body and its companion, and $v$ is the relative velocity between the two bodies. To extrapolate this to tertiary tides, it may be beneficial to imagine the moment when the inner binaries are in state $\alpha$ of Fig. 1 (left panel). Assuming that the tertiary is not rotating relative to the inner binary, and considering that $a_{\rm 2}{\gg}a_{\rm 1}$, one can see that, at this moment, $d$ should be substituted by $a_{\rm 2}$, and $v$ by the inner orbital velocity (the velocity at which $m_{\rm 1}$ and $m_{\rm 2}$ move relative to their COM), hereby denoted as $v_{\rm in}$. At moments when the triple system is not in state $\alpha$, the calculation of what values to substitute will be more problematic (possibly starting with the second term in Equation 12 of \citealt{1998MNRAS.300..292K}), and is beyond the scope of this paper. For the purposes of this study, we assume that $t_{\rm lag}$ is the same at all epochs, and hence we use $d=a_{\rm 2}$, $v=v_{\rm in}$ for all epochs, which is equivalent to
\begin{equation}
P=\frac{2{\pi}a_{\rm 2}}{v_{\rm in}}.
\label{perdef}
\end{equation}

The $\omega$ in Eq. \ref{lag_angle} is a measure of the lack of synchronism between the rotation of the perturbed body and its companion's orbit, and is therefore a function of the periodical variation of the tidal force acting upon the perturbed body. Thus, for tertiary tides, we set
\begin{equation}
{\omega}=2{\pi}\left({\frac{1}{P_{\rm in}}}-{\frac{1}{P_{\rm rot}}}\right)=2{\pi}\left({\frac{1}{P_{\rm in}}}-{\frac{1}{P_{\rm out}}}\right){\approx}\frac{2{\pi}}{P_{\rm in}},
\label{omegadef}
\end{equation}
\noindent where $P_{\rm in}$ is the orbital period of the inner binary, $P_{\rm rot}$ is the rotational period of $m_{\rm 3}$, which we assume in our Stage 1 simulations to be equal to the outer orbital period $P_{\rm out}$ due to tidal locking. 

As for the dissipation timescale $t_{\rm diss}$, since the tidal period is equal to $P_{\rm in}$, which is consistently greater than the convective timescale $t_{\rm conv}$ for all cases of TTs we consider, $t_{\rm diss}$ needs to be calculated via a prescription for fast tides, for which we use Eq. \ref{tdiss1}, and therefore 
\begin{equation}
t_{\rm diss}=\frac{t_{\rm conv}^{2}}{P_{\rm in}},
\label{tdissdef}
\end{equation}
\noindent where $t_{\rm conv}$ can be calculated, as with any normal star, via Eq. \ref{tconv} above.


Having ascertained the value of $t_{\rm lag}$, we then proceed with three sets of simulations to calculate the rate at which TTs extract orbital energy. 

In the first set, we run a simulation where $m_{\rm 1}$, $m_{\rm 2}$ and $m_{\rm 3}$ are all treated as point masses by solving the following set of equations using 8th-order Runge-Kutta:

\begin{equation}
\frac{{\rm d}{\bm v}_{\rm i}}{{\rm d}t}=\frac{Gm_{\rm j}}{r_{\rm ij}^{\rm 3}}({\bm R}_{\rm j}-{\bm R}_{\rm i})+\frac{Gm_{\rm k}}{r_{\rm ik}^{\rm 3}}({\bm R}_{\rm k}-{\bm R}_{\rm i})
\end{equation}


\noindent where blackfont denotes vectors, ${\bm R}_{\rm i}$ is the position vector of $m_{\rm i}$, i=(1,2,3), j=(2,1,1), and k=(3,3,2). This is done to check that the triple system is dynamically stable without the effects of tidal forces, and also serves to establish a baseline for the errors incurred during the simulations. 

In the second set, we run a 3-body simulation as before, but modulate the gravitational field of $m_{\rm 3}$ according to Eq. \ref{phi} with a giant tertiary (large radius), and a tidal lag. The lag is implemented by letting ${\zeta}$ at each timestep be what its equilibrium value would have been $t_{\rm lag}$ ago. In other words, 
\begin{equation}
\begin{split}
\frac{{\rm d}{\bm v}_{\rm 1}}{{\rm d}t}&=\frac{Gm_{\rm 2}}{r_{\rm 12}^{\rm 3}}({\bm R}_{\rm 2}-{\bm R}_{\rm 1})-\frac{V\left(r_{\rm 13},{\psi}_{\rm 1}\right)}{r_{\rm 13}^{\rm 2}}({\bm R}_{\rm 3}-{\bm R}_{\rm 1}),\\
\frac{{\rm d}{\bm v}_{\rm 2}}{{\rm d}t}&=\frac{Gm_{\rm 1}}{r_{\rm 21}^{\rm 3}}({\bm R}_{\rm 1}-{\bm R}_{\rm 2})-\frac{V\left(r_{\rm 23},{\psi}_{\rm 2}\right)}{r_{\rm 23}^{\rm 2}}({\bm R}_{\rm 3}-{\bm R}_{\rm 2}),\\
\frac{{\rm d}{\bm v}_{\rm 3}}{{\rm d}t}&=\frac{Gm_{\rm 1}}{r_{\rm 31}^{\rm 3}}({\bm R}_{\rm 1}-{\bm R}_{\rm 3})+\frac{Gm_{\rm 2}}{r_{\rm 32}^{\rm 3}}({\bm R}_{\rm 2}-{\bm R}_{\rm 3}).
\end{split}
\end{equation}
\noindent where the gravitational potential function $V\left(r,{\psi}\right)$ is given by Eq. \ref{phi}, and the value of $\zeta$ in the function is given as the equilibrium tide value $t_{\rm lag}$ ago. This is done to check that the triple system is dynamically stable after TTs are applied, and that the distortions of $m_{\rm 3}$ won't disintegrate the system before TTs come into effect. Theoretically, the energy extraction rate can be found using this method, but the errors induced by our approximations are larger than the benchmark set by our first set of simulations, and therefore we need another set of simulations to find this extraction rate. 

In the third set of simulations, we model the inner binary orbit only, adding an additional varying gravitational field centred at a distance $a_{\rm 2}$ from the COM of the inner binary. This field is equivalent to the effect of an $m_{\rm 3}$ tidally distorted by the orbiting inner binary, minus an $m_{\rm 3}$ tidally distorted by a point mass of $m_{\rm 1}+m_{\rm 2}$. The tidal lags are dealt with as before. The effective equations for this set of simulations are
\begin{equation}
\begin{split}
\frac{{\rm d}{\bm v}_{\rm 1}}{{\rm d}t}&=\frac{Gm_{\rm 2}}{r_{\rm 12}^{\rm 3}}({\bm R}_{\rm 2}-{\bm R}_{\rm 1})+k_{\rm 2}\left({\zeta}\left({\phi}\right)-{\zeta}_{\rm eq}\right)\\
&~~~~\left(\frac{R_{\rm 3}}{a_{\rm 2}}\right)^{2}P_{\rm 2}(\cos{\psi_{\rm 1}})\frac{Gm_{\rm 3}}{r_{\rm 13}^{\rm 3}}({\bm R}_{\rm 3}-{\bm R}_{\rm 1}),\\
\frac{{\rm d}{\bm v}_{\rm 2}}{{\rm d}t}&=\frac{Gm_{\rm 1}}{r_{\rm 21}^{\rm 3}}({\bm R}_{\rm 1}-{\bm R}_{\rm 2})+k_{\rm 2}\left({\zeta}\left({\phi}\right)-{\zeta}_{\rm eq}\right)\\
&~~~~\left(\frac{R_{\rm 3}}{a_{\rm 2}}\right)^{2}P_{\rm 2}(\cos{\psi_{\rm 2}})\frac{Gm_{\rm 3}}{r_{\rm 23}^{\rm 3}}({\bm R}_{\rm 3}-{\bm R}_{\rm 2})\\
{\bm v}_{\rm 3}&=\frac{{\bm v}_{\rm 1}+{\bm v}_{\rm 2}}{2}.
\end{split}
\end{equation}
\noindent here,
\begin{equation}
{\zeta}_{\rm eq}=\frac{(m_{\rm 1}+m_{\rm 2})}{m_{\rm 3}}\left(\frac{R_{\rm 3}}{a_{\rm 2}}\right)^3
\end{equation}
\noindent is the $\zeta$ value for an $m_{\rm 3}$ perturbed by a point mass of $m_{\rm 1}+m_{\rm 2}$ at the COM of the inner binary, and tidal lags are applied via $\zeta\left({\phi}\right)$ as in the second set. This excludes all effects other than TTs, and yields the rate of energy extraction, with which we then use to calibrate $\tau$ in our Stage 2 simulations (see below), by varying $\tau$ until the energy extraction rate matches that given by this model.


\subsection{Stage 2 Simulations}
\label{s1ss2}

While our Stage 1 simulations can provide the rate at which TTs extract energy from the inner binary, some details of the process (such as the rotation of the tertiary) are lost in the approximations. For a more convincing picture of how a hierarchical triple behaves under TTs, we resort to the following model.

Again, we consider the previous hierarchical coplanar triple system consisting of three stars with masses $m_{\rm 1}$, $m_{\rm 2}$ and $m_{\rm 3}$. 
As before, $m_{\rm 1}$ and $m_{\rm 2}$ are considered to be point masses, while
the tertiary is considered to be an oblate ellipsoid with mean radius $R_3$ and gravity field coefficients $J_2$, $C_{22}$ and $S_{22}$, sustained by the reference frame (${\bm I}$,${\bm J}$,${\bm K}$), where ${\bm K}$ is the axis of maximal inertia.
We furthermore assume that the spin axis of the tertiary, with rotation rate $\Omega$, is also along ${\bm K}$, and that ${\bm K}$ is orthogonal to the orbital plane (which corresponds to zero obliquity).
The gravitational potential of the tertiary is then given by \citep[e.g.][]{2013ApJ...767..128C}:
\begin{equation}
\begin{split}
V ({\bm r}) =& - \frac{G m_3}{r} - \frac{G m_3 R_{\rm 3}^2 J_2}{2 r^3} \\
& - \frac{3 G m_3 R_{\rm 3}^2}{r^3}  \big(  C_{22} \cos {2 \gamma}  - S_{22} \sin {2 \gamma} \big), 
\label{c1} 
\end{split}
\end{equation}
where
\begin{equation}
\cos {2 \gamma} =  ({\bm I} \cdot {\bm {\hat r}})^2 - ({\bm J} \cdot {\bm {\hat r}})^2
\quad \mathrm{and} \quad 
\sin {2 \gamma} = - 2 ({\bm I} \cdot {\bm {\hat r}}) ({\bm J} \cdot {\bm {\hat r}}) \ ,
\end{equation}
where ${\bm r}$ is a generic position with respect to the center of the tertiary, and ${\bm {\hat r}} = {\bm r} / r $ is the unit vector.
We neglect terms in $(R_{\rm 3}/r)^3$ (quadrupolar approximation).
We can also express ${\gamma} = \theta - f$, where $\theta$ is the rotation angle, and ${f}$ is the true longitude. 
The total potential energy of the system is thus given by
\begin{equation}
U ({\bm r}_1,{\bm r}_2) = - \frac{G m_1 m_2}{| {\bm r}_2-{\bm r}_1 |} + m_1 V({\bm r}_1) + m_2 V({\bm r}_2) \ , \label{c3} 
\end{equation}
where ${\bm r}_i = {\bm R}_i - {\bm R}_3$, and ${\bm R}_i$ is the position of the star with mass $m_i$ in an inertial frame.
Note that the quantities $a_1$ and $a_2$ (Jacobi coordinates) 
are not the norms of ${\bm r}_1$ and ${\bm r}_2$, which are astrocentric coordinates (see appendix \ref{delayed_to_visco} for more details).

The equations of motion governing the orbital evolution of the system in an inertial frame are given by: 
\begin{equation}
\frac{d^2 {\bm R}_i}{dt^2} = - \frac{1}{m_i} \frac{\partial U}{\partial {\bm R}_i} = - \frac{1}{m_i} \frac{\partial U}{\partial {\bm r}_i} \ , \label{161223b} 
\end{equation}
\begin{equation}
\frac{d^2 {\bm R}_3}{dt^2} = - \frac{1}{m_3} \frac{\partial U}{\partial {\bm R}_3} = \frac{1}{m_3} \left( \frac{\partial U}{\partial {\bm r}_1} + \frac{\partial U}{\partial {\bm r}_2} \right) \ , \label{161223c} 
\end{equation}
with
\begin{eqnarray}
\frac{\partial U}{\partial {\bm r}_i} \!&=&\! (-1)^i \frac{G m_1 m_2}{|{\bm r}_2-{\bm r}_1|^3} ({\bm r}_2-{\bm r}_1) 
+ \frac{G m_i m_3}{r_i^3} {\bm r}_i \nonumber \\ && 
+ \frac{3 G m_i m_3 R_{\rm 3}^2}{2 r_i^5} \left[ J_2 + 6 \big(  C_{22} \cos {2 \gamma}_i  - S_{22} \sin {2 \gamma}_i \big) \right]  {\bm r}_i \nonumber \\ && 
- \frac{6 G m_i m_3 R_{\rm 3}^2}{r_i^5}  \big( C_{22} \sin {2 \gamma}_i  + S_{22} \cos {2 \gamma}_i \big) {\bm K} \times {\bm r}_i \ . 
\label{161223d} 
\end{eqnarray}

In an astrocentric frame they simply become
\begin{equation}
\frac{d^2 {\bm r}_i}{dt^2} = - \left( \frac{1}{m_i} + \frac{1}{m_3} \right) \frac{\partial U}{\partial {\bm r}_i} - \frac{1}{m_3} \frac{\partial U}{\partial {\bm r}_j}  \ , \label{161223e}
\end{equation}
where $i = 1,2$ and $j=3-i$.


The torque acting to modify the rotation of the tertiary is given by
\begin{equation}
I_{\rm 3} \frac{d {\Omega}}{dt} = \left( {\bm r}_1 \times \frac{\partial U}{\partial {\bm r}_1} + {\bm r}_2 \times \frac{\partial U}{\partial {\bm r}_2} \right) \cdot {\bm K}  \ , \label{161223f} 
\end{equation}
for a tertiary of constant radius, where $I_{\rm 3}$ is the principal moment of inertia of $m_{\rm 3}$ along the axis ${\bm K}$.
We hence obtain for the rotation angle $\dot \theta = {\Omega}$:
\begin{eqnarray} 
\frac{d^2 \theta}{dt^2} \!&=&\! -\frac{6Gm_1m_3R_{\rm 3}^2}{I_{\rm 3}r_1^3} \left[ C_{22}\sin2{\gamma}_1 + S_{22}\cos2{\gamma}_1\right] \nonumber \\ && 
-\frac{6Gm_2m_3R_{\rm 3}^2}{I_{\rm 3}r_2^3} \left[ C_{22}\sin2{\gamma}_2 + S_{22}\cos2{\gamma}_2\right] \ .
\label{161223g} 
\end{eqnarray}
\noindent For a tertiary with varying radius, the above equation becomes
\begin{eqnarray} 
\frac{d^2 \theta}{dt^2} \!&=&\! -2{\Omega}\frac{\dot R_{\rm 3}}{R_{\rm 3}}-\frac{6Gm_1m_3R_{\rm 3}^2}{I_{\rm 3}r_1^3} \left[ C_{22}\sin2{\gamma}_1 + S_{22}\cos2{\gamma}_1\right] \nonumber \\ && 
-\frac{6Gm_2m_3R_{\rm 3}^2}{I_{\rm 3}r_2^3} \left[ C_{22}\sin2{\gamma}_2 + S_{22}\cos2{\gamma}_2\right] \ .
\end{eqnarray}
\noindent When this is the case, we find a discrete $R_{\rm 3}=R_{\rm 3}(t)$ via stellar evolution codes, and use cubic spline interpolation to determine both $R_{\rm 3}$ and ${\rm d}R_{\rm 3}/{\rm d}t$ at each epoch.

The tertiary is deformed under the action of self rotation and tides. 
Therefore, the gravity field coefficients can change with time
as the shape of the tertiary is continuously adapting to the equilibrium figure. 
According to the Maxwell viscoelastic rheology, the deformation law for these coefficients is given by \citep[e.g.][]{2014A&A...571A..50C}:
\begin{eqnarray}\label{max1}
&&J_2+\tau\dot{J}_2 =  J_2^{r} + J_2^t \ ,\nonumber\\
&&C_{22}+\tau\dot{C}_{22}  =  C_{22}^t \ ,\\
&&S_{22}+\tau\dot{S}_{22} = S_{22}^t \ ,\nonumber
\end{eqnarray}
where
\begin{equation}\label{j2r}
J_2^{r} = k_2 \frac{{\Omega}^2R_{\rm 3}^3}{3Gm_3}
\end{equation}
is the rotational deformation, and 
\begin{eqnarray}\label{max2}
&&J_2^t=k_2 \frac{m_1}{2m_3}\left(\frac{R_{\rm 3}}{r_1}\right)^3 + k_2 \frac{m_2}{2m_3}\left(\frac{R_{\rm 3}}{r_2}\right)^3 \ ,\\
&&C_{22}^t=\frac{k_2}{4}\frac{m_1}{m_3}\left(\frac{R_{\rm 3}}{r_1}\right)^3\cos2{\gamma}_1 + \frac{k_2}{4}\frac{m_2}{m_3}\left(\frac{R_{\rm 3}}{r_2}\right)^3\cos2{\gamma}_2 \ , \nonumber \\
&&S_{22}^t=-\frac{k_2}{4}\frac{m_1}{m_3}\left(\frac{R_{\rm 3}}{r_1}\right)^3\sin2{\gamma}_1 -\frac{k_2}{4}\frac{m_2}{m_3}\left(\frac{R_{\rm 3}}{r_2}\right)^3\sin2{\gamma}_2 \nonumber \ ,
\end{eqnarray}
are the tidal equilibrium values for the gravity coefficients,
and $\tau$ is the relaxation time of the tertiary in response to deformation. Usually, $\tau = \tau_v + \tau_e$, where $\tau_v$ and $\tau_e$ are the viscous (or fluid) and Maxwell (or elastic) relaxation times, respectively. However, for simplicity, we consider $\tau_e=0$, since this term does not contribute to the tidal dissipation \citep[see][]{2014A&A...571A..50C}. This $\tau$ is the previously mentioned unknown parameter calibrated using our Stage 1 simulations. For an evolving tertiary, $\tau$ admittedly changes with time, but its degree of variation is not prominent enough to warrant treating it as a variable, for the purposes of the simulations mentioned in this paper.

\section{Examples of Systems Undergoing TTs}

To showcase the effects of TTs, as well as the capabilities of our simulations, we run two sets of simulations: one for a purely hypothetical system consisting of two WDs and a MS star, with orbital parameters designed to maximise TTs, and the other for an observed multiple star system, namely HD97131. This section provides the details of these systems, as well as our results.

\subsection{Hypothetical Scenario}
\label{s2ss1}

Here, we consider a purely hypothetical hierarchical triple, in which the inner binary consists of a pair of tidally locked white dwarfs (WDs), and the tertiary is a MS star, tidally locked to the inner binary's COM. The masses are given as $m_{\rm 1}=m_{\rm 2}=0.8M_{\odot}$, $m_{\rm 3}=1.6M_{\odot}$, and the orbital semimajor axes as $a_{\rm 1}=0.2$AU, $a_{\rm 2}=2$AU. The orbits are set to be coplanar and prograde, and all orbits are given to be circular. In this system, the WDs can readily be approximated as point masses, thus forming a ripe testing ground for tertiary tides. It should be noted that its circular and coplanar orbits also preclude Lidov-Kozai Resonance \citep{1962AJ.....67..591K,1962P&SS....9..719L} from this system. 

There are two main reasons why we choose such a system for our demonstration. The first is that this system is realistic - with an inner orbit of 25.8 days and an outer orbit of 577.4 days, this system has similar orbital periods to triple systems that have actually been observed. In fact, extensive studies by \citet{1997A&AS..124...75T} have found many triple systems with inner and outer orbital periods close to and straddling these (see their Figure 3). The second is that this system is stable according to conventional wisdom, if all three bodies were point masses. Adopting the methods and criteria of \citet{2005A&A...434..355M}, we check this by following the dynamical evolution of the system over 4000 outer orbits using RK8, and examining their trajectories. The orbits are found to be stable, which is expected, given that the system falls within well-established stable zones \citep[e.g.][]{2005A&A...434..355M,2014ApJ...780...14C}.

Using our two-stage simulation method, we find that the effect of TTs is negligible when $m_{\rm 3}$ is still an MS star. This is expected, since the radius of a $1.6M_{\odot}$ MS star is relatively small (about 1$R_{\odot}$), whereas tidal phenomena typically require radii on the order of the Roche Lobes of the systems involved. However, MS stars evolve into red giants later in their lifetimes, and red giants have much larger radii. Using well-established stellar evolution algorithms \citep{1973MNRAS.163..279E,1995MNRAS.274..964P,2011ApJS..192....3P}, we find that a $1.6M_{\odot}$ star stays in the red giant phase for many Myrs, during which its radius expands to more than 140 solar radii. This radius is close to, but just short of, its Roche Lobe, and therefore we need not consider the effects of Roche Lobe overflow.

\begin{figure}
\includegraphics[scale=0.33, angle=270]{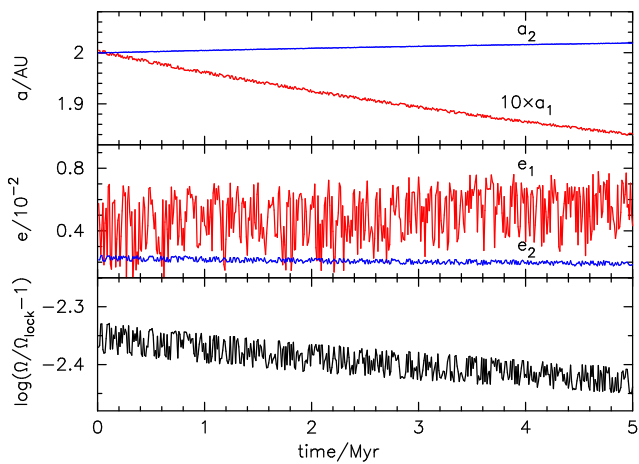}
\caption{Orbital evolution of a hierarchical triple with $m_{\rm 1}=m_{\rm 2}=0.8M_{\odot}$, $m_{\rm 3}=1.6M_{\odot}$, $a_{\rm 1}=0.2$AU, $a_{\rm 2}=2$AU, $e_{\rm 1}=e_{\rm 2}=0$, and a constant tertiary radius of 100 solar radii. The inner binary orbit shrinks significantly within just a few Myrs due to TTs alone, while other orbital parameters also undergo some evolution. The rotational velocity of $m_{\rm 3}$ deviates from being perfectly tidally locked with the inner binary by a small amount, due to reasons explained in the text. \label{Fig3}}
\end{figure}

Again adopting our two-stage simulation method, and assuming a constant radius of 100 solar radii for $m_{\rm 3}$, we retrieve a $\tau$ of 0.534 years (see Table \ref{sim_params}), and find that the inner binary orbit shrinks significantly within just a few Myrs due to TTs alone (Fig. \ref{Fig3}). Throughout the inner binary orbital shrinkage, angular momentum from the inner orbit is transferred to the outer orbit, and $a_{\rm 2}$ marginally increases as a result, though not enough to shut down further shrinkage of $a_{\rm 1}$ due to TTs.

\begin{table}
	\centering
	\caption{Initial parameters for our second-stage simulations in the tertiary RGB phase for both our hypothetical scenario and HD97131.}
	\label{sim_params}
	\begin{tabular}{ccc}
		\hline
		Parameter & Hypothetical Scenario & HD97131 \\
		\hline
                $a_1$/AU & 0.2 & 0.0373 \\
                $a_2$/AU & 2.0 & 0.7955 \\
                $e_1$ & 0 & 0 \\
                $e_2$ & 0 & 0.191 \\
                $m_1$/$M_{\odot}$ & 0.8 & 1.29 \\
                $m_2$/$M_{\odot}$ & 0.8 & 0.90 \\
                $m_3$/$M_{\odot}$ & 1.6 & 1.50 \\
		$\tau$/years & 0.534 & 0.019 \\
		\hline
	\end{tabular}
\end{table}

We also find that, after $m_{\rm 3}$ becomes a red giant, its rotational velocity is never exactly locked to the inner binary, even though the deviation is small. This is probably due to the fact that, for a perfectly locked $m_{\rm 3}$, the mass elements of $m_{\rm 3}$ that are closer to the inner binary will have a tendency to move in the same direction as the closer inner binary component, thereby inducing a rotation that deviates from a perfectly tidally locked scenario. While this deviation is unlikely to be of much physical significance in our model, calculations pertaining to tidal effects in close triple systems, performed with models of dynamical oscillation modes under the assumption of perfect tidal locking, may require special attention in this regard.

\begin{figure}
\includegraphics[scale=0.34, angle=270]{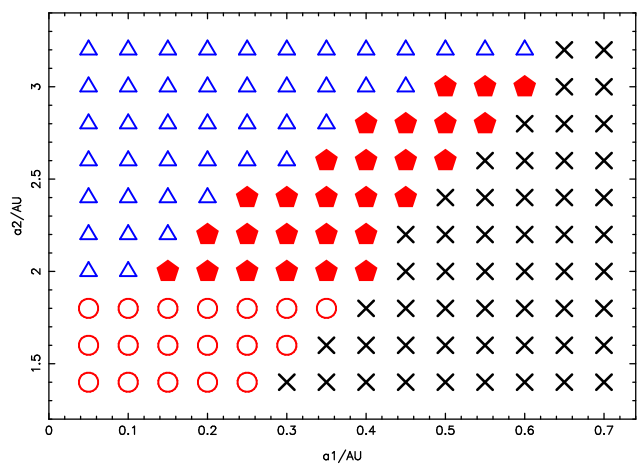}
\caption{Prominence of tertiary tides in $a_{\rm 1}$-$a_{\rm 2}$ space for our hypothetical hierarchical triple system with $m_{\rm 1}=m_{\rm 2}=0.8M_{\odot}$, $m_{\rm 3}=1.6M_{\odot}$, all orbits being coplanar and circular. The black crosses indicate the region in which $a_{\rm 2}/a_{\rm 1}$ is too small, and the system is dynamically unstable; the blue triangles cover the region in which $a_{\rm 2}$ is too large, and TTs have no noticeable effect; the red circles represent areas where $m_{\rm 3}$ would fill its Roche lobe, in which Roche lobe overflow will compete with TTs for dominance. Only in the region with the filled red pentagons are TTs the exclusive dominating factor in merging the binary. \label{Fig4}}
\end{figure}

But what if the inner or outer orbital separations in the triple system were larger or smaller? After all, realistic triple systems have a great range of values for $a_{\rm 1}$ and $a_{\rm 2}$. To check the separation dependence of TTs, we conduct a grid of first-stage simulations in $a_{\rm 1}$-$a_{\rm 2}$ space for the same $m_{\rm 1}=m_{\rm 2}=0.8M_{\odot}$, $m_{\rm 3}=1.6M_{\odot}$ system, and check how fast TTs can remove orbital energy from the inner binary for each set of ($a_{\rm 1}$, $a_{\rm 2}$). The conclusion is that, for different sets of ($a_{\rm 1}$, $a_{\rm 2}$), one of 4 different scenarios are possible: (i) if $a_{\rm 2}/a_{\rm 1}$ is too small, the system is dynamically unstable, and the orbits will evolve unpredictably whether TTs are considered or not; (ii) if $a_{\rm 2}$ is too large, TTs will have no noticeable effect; (iii) if $a_{\rm 2}$ is too small, $m_{\rm 3}$ will fill its Roche Lobe at some point during its evolution. While this does not invalidate the influence of TTs (TTs can lead to very significant orbital shrinkage of the inner binary before Roche Lobe overflow even begins, as demonstrated later in this section), it does lead to complications as to which effect dominates the evolution of the binary thereafter, which are beyond the scope of this paper; (iv) only in a triangular region straddled by these three regions are TTs the exclusive dominating factor. We plot these four regions in $a_{\rm 1}$-$a_{\rm 2}$ space for our $m_{\rm 1}=m_{\rm 2}=0.8M_{\odot}$, $m_{\rm 3}=1.6M_{\odot}$ system (Fig. \ref{Fig4}). It can be seen that it is only in some of the closest hierarchical triples that TTs play a dominant role.


\subsection{HD97131}
\label{s2ss2}

How would TTs influence a realistic hierarchical triple system? To answer this question, we refer ourselves to the real-world hierarchical triple HD97131. HD97131 is a coplanar triple system \citep{2003AJ....125..825T} with $m_{\rm 3}$ being an MS star of spectral type F0. The inner orbit (between $m_{\rm 1}$ and $m_{\rm 2}$) is circular, while the outer orbit (which we assume to be prograde) has an eccentricity of $e_{\rm 2}=0.191$. The other relevant orbital parameters are $m_{\rm 1}=1.29M_{\odot}$, $m_{\rm 2}=0.90M_{\odot}$, $m_{\rm 3}=1.50M_{\odot}$, $a_{\rm 1}=0.0373$AU, and $a_{\rm 2}=0.7955$AU \citep{1997A&AS..124...75T,2010yCat..73890925T}. At such a small $a_{\rm 2}$, the Roche Lobe for $m_{\rm 3}$ is small, only 57.1 solar radii \citep{1983ApJ...268..368E}. Thus, $m_{\rm 3}$ will inevitably fill its Roche Lobe during its red giant phase. However, since the effects of Roche Lobe mass transfer do not become significant until after its onset, this fact will not affect our analysis of TTs, which will shrink $a_{\rm 1}$ long before this happens. Unlike our previous simulation, we account for the radius evolution of $m_{\rm 3}$ by calculating the radius as a function of time using the aforementioned stellar evolution codes \citep{1973MNRAS.163..279E,1995MNRAS.274..964P,2011ApJS..192....3P}, and performing a cubic spline interpolation on the results. For our following simulations, we use the final 4.8 Myrs of the radius evolution of $m_{\rm 3}$ up to 1 Myr after Roche Lobe overflow. This means that our simulation starts with the initial orbital parameters, but with $m_{\rm 3}$ already well into its RGB phase, and filling its Roche Lobe at $t=$3.8 Myrs.

To simulate HD97131, we use our 2-stage method described in \S 2. However, since many of the assumptions regarding our first-stage simulations break down for systems with $m_{\rm 1}{\neq}m_{\rm 2}$ and $e_{\rm 2}{\neq}0$, we make the following modifications to our methods. For our first-stage simulations, we set the masses of both inner binary components to $\frac{m_{\rm 1}+m_{\rm 2}}{2}$, and set the tertiary in a circular orbit with a semimajor axis of $a_{\rm 2}(1-e_{\rm 2}^{2})$. The justification for the latter is that, assuming conservation of angular momentum, the semimajor axis of the outer orbit will evolve to that particular value if the orbit were to be tidally circularised. Our results show that this is indeed the case. With these modifications to the system, our original assumptions hold, and the first-stage simulations can be conducted. While this leads to a first stage simulation of a system somewhat different from the actual HD97131, this difference is not important, as our first stage simulations are only used to calibrate the value of $\tau$ for our second-stage simulations, which are responsible for the recovery of exact details of the orbital evolution. We find a $\tau$ of 0.019 years. For the second-stage simulations, we use the orbital parameters of the actual HD97131 (as documented in Table \ref{sim_params}).

\begin{figure}
\includegraphics[scale=0.33, angle=270]{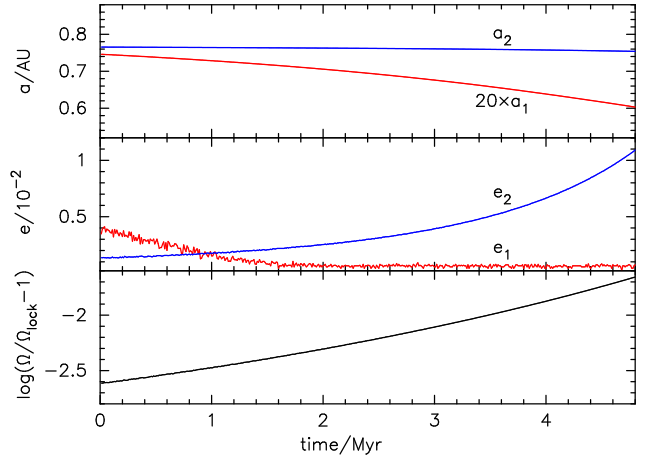}
\caption{Projected orbital evolution for HD97131, after its tertiary becomes a red giant. It can be seen that the inner orbital separation $a_{\rm 1}$ decreases significantly due to TTs, while the evolution of the outer orbital separation $a_{\rm 2}$ is negligible. The evolution of the orbital eccentricities is evident, as is that of $\Omega$, the rotation of $m_{\rm 3}$. Note that the initial $e_{\rm 2}=0.191$ vanishes in just a few thousand years, and is consequently not visible in this plot. $\Omega$ evolves to deviate from the tidally locked value as expected, due to reasons explained in the paper. \label{Fig5}}
\end{figure}

Tracing the orbital evolution of HD97131 during the red giant phase of $m_{\rm 3}$, we find that the outer orbit is rapidly circularised, reducing $e_{\rm 2}$ to less than $0.01$ in just a few thousand years. This is expected \citep[e.g.][]{2016CeMDA.126..189C}, and will therefore warrant no further attention here. Thereafter, the inner binary orbit shrinks as witnessed in our previous simulations, with some minor evolution of the other orbital parameters (Fig. \ref{Fig5}). It should be noted that the GW merging timescale of the inner binary is brought down to less than half its original value during this process \citep[see][]{1964PhRv..136.1224P}. The slight deviation from exact tidal locking in the rotation of $m_{\rm 3}$ is again seen.

\section{Discussion}

Our results unequivocally show that TTs have a profound impact on very close hierarchical triples. While it is evident that this translates into a negligible effect when considering stellar populations in general, our understanding of certain exotic systems can be spurious if it were to be neglected altogether.

For starters, gravitational wave mergers \citep[e.g.][]{2016PhRvL.116f1102A} require very close massive objects as progenitors. It is also well known that multiplicity is enhanced in stellar objects of such masses \citep[e.g.][]{2012Sci...337..444S}, and that GW mergers can arise from multiple interactions in globular clusters \citep[e.g.][]{2016ApJ...824L...8R}. In both of these cases, TTs will be much more prevalent than in any general stellar population, though it is difficult to be certain by how much.

Of the many possible sources \citep[e.g.][]{1973ApJ...186.1007W,2004MNRAS.350.1301H,1984ApJ...277..355W,2013ApJ...770L...8P} of Type Ia supernovae (SNe Ia for short), one proposed progenitor system involving three-body interactions has received a certain degree of attention in recent years (\citealt{2011ApJ...741...82T},\citealt{2013ApJ...766...64S}, see also \citealt{2007ApJ...669.1298F} and \citealt{2015MNRAS.454L..61D}), despite the fact that it is unlikely to be one of the main sources of SN Ia production \citep{2013MNRAS.430.2262H}. In these systems, the existence of a tertiary drives a WD binary into a merger or collision, by means of Lidov-Kozai oscillations. While Lidov-Kozai oscillations are less diminished by large values of $a_{\rm 2}$ than are TTs, it is conceivable that such systems preferably have small values of $a_{\rm 2}$, and therefore at least some of these systems must be susceptible to TTs. Furthermore, Lidov-Kozai oscillations are only an issue when mutual inclinations between the inner and outer orbits are high ($\sim40$ degrees or more), whereas TTs work for both coplanar and highly inclined systems. Thus, SN Ia production rates from such progenitor systems will be underestimated, should TTs be left unaccounted for. Another analogous issue is the enrichment in the high-mass end of WD mass functions \citep[e.g.][]{2015MNRAS.452.1637R}, which cannot be explained by WD mergers via gravitational waves alone. While TTs are unlikely to have contributed significantly to this enrichment, they could potentially amplify the rate of WD mergers if WD binaries are found to have a greater degree of multiplicity than previously thought.

Last but not least, there have been attempts to explain observational phenomena with models of binary mergers occurring inside the envelopes of giant stars \citep[e.g.][]{2017MNRAS.471.3456H}. Should such studies ever reach the point where a detailed simulation of a progenitor system is required, TTs must be considered, as any binary must undergo a phase of non-negligible TTs before it can end up inside the envelope of a giant star.

In summary, TTs should play a pivotal role in the orbital evolution of certain systems. This role is even more preponderant when one considers the fact that a smaller $a_{\rm 1}$ can further exacerbate other mechanisms that drive the inner binary closer together (e.g. gravitational waves). The only limiting factor of their general influence on 3-body evolution is the fraction of systems that will experience significant TTs; as of yet, observational evidence of how frequently they occur is not available. An examination of observed triple systems \citep{1997A&AS..124...75T,2010yCat..73890925T} seems to imply that only a very small fraction (${\sim}1\%$) will undergo significant TTs in the future; however, since TTs and other effects that act in close triples have a tendency to destroy their host systems, resulting in them ending up as binaries, not to mention observational biases that may limit the amount of very close triples seen, it is fairly hard to say what fraction of triples would be influenced by future TTs at the time of their birth. Perhaps the best way to ascertain this would be to collect samples of hierarchical triples in which the tertiaries are stars that have already evolved beyond their red giant phases, and to compare their $a_{\rm 1}$ values against those of triples with less advanced tertiaries; such observations of post-red-giant tertiaries, however, are currently rare. Opportunities to directly observe TTs in action may present themselves from time to time, judging from the existence of systems such as HD181068 \citep{2011Sci...332..216D} and KOI-126 \citep{2011ApJ...740L..25F}, and theoretical modelling by means of adding TTs to existing triple star evolution codes \citep[e.g.][]{2017AAS...22932605T} may also shed further light on this phenomenon that we know so little about, but such endeavours will be the contents of a future paper.

\section*{Acknowledgements}

We thank a number of colleagues, including but not limited to Sverre Aarseth, Robert Izzard, Simon Jefferey, Philipp Podsiadlowski, Bo Wang, You Wu, Cong Yu, and Jilin Zhou, for discussion and encouragement.

This work was partly supported by the 
Natural Science Foundation of China (Grant Nos. 11521303, 11733008, 11390374), 
the Science and Technology Innovation Talent Programme of Yunnan Province (Grant No. 2017HC018), 
the Chinese Academy of Sciences (Grant No. KJZD-EW-M06-01),
CIDMA strategic project (UID/MAT/04106/2013), 
ENGAGE SKA (POCI-01-0145-FEDER-022217), and 
PHOBOS (POCI-01-0145-FEDER-029932),
funded by COMPETE 2020 and FCT, Portugal.








\appendix

\section{Theoretical Calculations of TTs under Ideal Conditions}
\label{app_other}

For a hierarchical triple undergoing TTs, if it were the case that the internal dissipation of $m_{\rm 3}$ is infinitely efficient (which it is not), all the energy stored in the difference in the self gravitational potential energy of $m_{\rm 3}$ between states $\alpha$ and $\beta$ (see Fig. \ref{Fig1}) would be effectively dissipated. Therefore, the amount of energy extracted from the inner binary orbit between $m_{\rm 1}$ and $m_{\rm 2}$, during the time it takes for the system to evolve from state $\alpha$ to state $\beta$, must be the same as the self gravitational potential energy difference of $m_{\rm 3}$. As the system evolves back and forth twice between states $\alpha$ and $\beta$ for every inner orbit, the rate of energy extraction from the inner orbit must equal four times this energy difference per inner orbital period.

How, then, should one calculate the difference in self gravitational potential energy between the third body at states $\alpha$ and $\beta$? A spherical, perfectly homogeneous elastic body under the influence of a tidal force will assume the geometric shape of an ellipsoid. The self-gravitational potential energy of a homogeneous triaxial ellipsoid can be calculated from the equations given in \citet{2000astro.ph..3064S}, repeated below:
\begin{equation}
\begin{split}
E_{\rm P}&={\frac{3}{10}}GM^{2}{\int_{0}^{+\infty}}{\frac{{\rm d}s}{Q_{\rm s}}},\\
Q_{\rm s}&=\sqrt{({a_{\rm x}}^{2}+s)({a_{\rm y}}^{2}+s)({a_{\rm z}}^{2}+s)}.
\end{split}
\label{q1}
\end{equation}
\noindent Here, $W$ is the potential energy, $a_{\rm x}$, $a_{\rm y}$, and $a_{\rm z}$ are the semi-axes of the ellipsoid along the $x$, $y$ and $z$ axes, respectively, and $M$ is the mass of the body. Thus, the potential energy difference between states $\alpha$ and $\beta$ is simply
\begin{equation}
{\Delta}E_{\rm P}={\frac{3}{10}}GM^{2}{\int_{0}^{+\infty}}({\frac{1}{Q_{\rm \alpha}}}-{\frac{1}{Q_{\rm \beta}}}){\rm d}s.
\label{q2}
\end{equation}
\noindent However, this treatment has the inconvenience that it is difficult to modify for an inhomogeneous ellipsoid, which is something we have to address later in this section. Therefore, we adopt a different approach, as follows.

Under the influence of a small geometrical distortion, which is true in our case, the ellipsoid resulting from the aforementioned homogeneous spherical body is still roughly spherical in shape. Adopting a spherical coordinate system ($r$,$\psi$,$\xi$) where $\psi=0$ along the direction pointing towards the COM of the inner binary, the difference of the self-gravitational potential energy, between the initial sphere and that of the ellipsoid resulting from tidal influence, is simply the potential energy difference due to the change in radius at every ($\psi$,$\xi$), integrated over the surface of the sphere. Further assuming that the density difference of the body before and after applying the tidal force is negligible, the gravitational potential energy difference at ($\psi$,$\xi$) is equal to $(|{\Delta}R(\psi,\xi)|{\rho}g){\times}({\frac{1}{2}}|{\Delta}R(\psi,\xi)|)$, and the integral of this over the entire surface of the sphere is 
\begin{equation}
\begin{split}
E_{\rm P,ell}-E_{\rm P,sph}&=\int_{0}^{2\pi}\int_{0}^{\pi}(|{\Delta}R(\psi,\xi)|{\rho}g)\\
&({\frac{1}{2}}|{\Delta}R(\psi,\xi)|)R_{\rm 3}^{2}\sin{\psi}{\rm d}{\psi}{\rm d}{\xi}.
\end{split}
\label{deltap1}
\end{equation}
\noindent Here, $E_{\rm P,ell}$ and $E_{\rm P,sph}$ are the potential energies of the body when it is a homogeneous ellipsoid and a homogeneous sphere, respectively; $|{\Delta}R(\psi,\xi)|$ is the absolute value of the change in radius at ($\psi$,$\xi$), $(|{\Delta}R(\psi,\xi)|{\rho}g)$is the amount of mass displaced at ($\psi$,$\xi$) due to the change in radius, $({\frac{1}{2}}|{\Delta}R(\psi,\xi)|)$ is the displacement of the centre of mass of the displaced mass, $R_{\rm 3}$ is the radius of the original sphere, and $R_{3}^{2}\sin{\psi}$ is the Jacobian determinant for spherical integration. The general expression for ${\Delta}R(\psi,\xi)$ can be derived from
\begin{equation}
\begin{split}
{\Delta}R(\psi,\xi)&=R(\psi,\xi)-R_{\rm 3},\\
R(\psi,\xi)&=R_{\rm 3}[1+\frac{5}{3}k_{\rm 2}{\zeta}P_{\rm 2}(\cos{\psi})],
\end{split}
\label{deltar1}
\end{equation}
\noindent where $k_{\rm 2}$ is the Love number, which is equal to $\frac{3}{2}$ for a homogeneous fluid body, ${\zeta}$ is a parameter reflecting the magnitude of the tidal effects, the value of which we will deal with later in this section, and $P_{\rm 2}(\cos{\psi})$ is a Legendre polynomial, equal to ${\frac{1}{2}}(3{\cos}^2{\psi}-1)$. Since all stars are fluid bodies, we set $k_{\rm 2}=\frac{3}{2}$ for a homogeneous body, and since $R({\psi},{\xi})$ does not explicitly contain $\xi$, Eq. \ref{deltar1} thus becomes
\begin{equation}
\begin{split}
{\Delta}R({\psi},{\xi})&={\Delta}R(\psi)\\
&={\frac{5}{4}}R_{\rm 3}{\zeta}(3{\cos}^2{\psi}-1),
\end{split}
\label{delta_R}
\end{equation}
\noindent and, by extension, Eq. \ref{deltap1} can be calculated, by substituting the expressions for ${\rho}$ and $g$, as well as Eq. \ref{delta_R}, to be
\begin{equation}
\begin{split}
E_{\rm P,ell}-E_{\rm P,sph}&=\int_{0}^{2\pi}\int_{0}^{\pi}{\frac{1}{2}}{\rho}g({\Delta}R(\psi))^2{R_{3}}^{2}\sin{\psi}{\rm d}{\psi}{\rm d}{\xi}\\
&=\int_{0}^{2\pi}\int_{0}^{\pi}{\frac{1}{2}}(\frac{m_{\rm 3}}{\frac{4}{3}{\pi}{R_{3}}^{3}})(\frac{Gm_{\rm 3}}{{R_{3}}^{2}})\\
&~~~~({\frac{5}{4}}R_{3}{\zeta}(3{\cos}^2{\psi}-1))^2{R_{3}}^{2}\sin{\psi}{\rm d}{\psi}{\rm d}{\xi}\\
&={\frac{75}{128{\pi}}}{\frac{Gm_{\rm 3}^2{\zeta}^2}{R_{3}}}\int_{0}^{2\pi}\int_{0}^{\pi}(3{\cos}^2{\psi}-1)^2\\
&~~~~\sin{\psi}{\rm d}{\psi}{\rm d}{\xi}.
\end{split}
\end{equation}
\noindent Since
\begin{equation}
\begin{split}
&~~~~{\int}(3{\cos}^{2}x-1)^2{\sin}x{\rm d}x\\
&=-\frac{9}{5}{\cos}^{5}x+2{\cos}^{3}x-{\cos}x,
\end{split}
\end{equation}
\noindent the previous equation becomes
\begin{equation}
\begin{split}
E_{\rm P,ell}-E_{\rm P,sph}&={\frac{75}{128{\pi}}}{\frac{Gm_{\rm 3}^2{\zeta}^2}{R_{3}}}\int_{0}^{2\pi}\int_{0}^{\pi}(3{\cos}^2{\psi}-1)^2\\
&~~~~\sin{\psi}{\rm d}{\psi}{\rm d}{\xi}\\
&={\frac{75}{128{\pi}}}{\frac{Gm_{\rm 3}^2{\zeta}^2}{R_{3}}}({2\pi})\frac{8}{5}\\
&={\frac{15}{8}}{\frac{Gm_{\rm 3}^2{\zeta}^2}{R_{3}}},
\end{split}
\label{deltap2}
\end{equation}
\noindent where $G$ is the gravitational constant.

For a two-body tide, where one object experiences tidal force from the other, ${\zeta}$ is given by
\begin{equation}
{\zeta}=\frac{m_{\rm per}}{M}\left(\frac{R_{\rm 0}}{a}\right)^3
\label{zeta1}
\end{equation}
\noindent where $m_{\rm per}$ is the mass of the perturbing body, $M$ is the mass of the perturbed body, $R_{\rm 0}$ is the spherical radius of the receiving body in the absence of tidal forces, and $a$ is the distance between the two bodies. It follows that, in our situation, when $m=m_{\rm 1}=m_{\rm 2}$, for states $\alpha$ and $\beta$ (see Figure \ref{Fig1} for definition),
\begin{equation}
\begin{split}
{\zeta}_{\rm \alpha}&=\left(\frac{a_{\rm 2}^3}{(a_{\rm 2}+\frac{1}{2}a_{\rm 1})^3}+\frac{a_{\rm 2}^3}{(a_{\rm 2}-\frac{1}{2}a_{\rm 1})^3}\right)\frac{m}{m_{\rm 3}}\left(\frac{R_{\rm 3}}{a_{\rm 2}}\right)^3\\
{\zeta}_{\rm \beta}&=\left[3\left(\frac{a_{\rm 2}}{\sqrt{a_{\rm 2}^2+(\frac{1}{2}a_{\rm 1})^2}}\right)^5-\left(\frac{a_{\rm 2}}{\sqrt{a_{\rm 2}^2+(\frac{1}{2}a_{\rm 1})^2}}\right)^3\right]\frac{m}{m_{\rm 3}}\left(\frac{R_{\rm 3}}{a_{\rm 2}}\right)^3,
\end{split}
\label{modzeta1}
\end{equation}
\noindent where $a_{\rm 1}$, $a_{\rm 2}$, $m_{\rm 1}$, $m$ and $m_{\rm 3}$ have already been defined in the text. Setting $u=({a_{\rm 1}}/{a_{\rm 2}})^2$, the above equations are strictly equivalent to
\begin{equation}
\begin{split}
{\zeta}_{\rm \alpha}&=\frac{m}{m_{\rm 3}}\left(\frac{R_{\rm 3}}{a_{\rm 2}}\right)^3\frac{2+\frac{3}{2}u}{\left( 1-\frac{1}{4}u\right)^3}\\
{\zeta}_{\rm \beta}&=\frac{m}{m_{\rm 3}}\left(\frac{R_{\rm 3}}{a_{\rm 2}}\right)^3\left[3\left(\frac{1}{1+\frac{1}{4}u}\right)^{5/2}-\left(\frac{1}{1+\frac{1}{4}u}\right)^{3/2}\right].
\end{split}
\end{equation}
\noindent When $a_{\rm 1}<<a_{\rm 2}$, it follows that $u$ is small, and therefore terms of order $u^2$ and higher can be omitted:
\begin{equation}
\begin{split}
\frac{2+\frac{3}{2}u}{\left( 1-\frac{1}{4}u\right)^3}&=\left(2+\frac{3}{2}u\right)\left(1+\left(\frac{3}{4}u-\frac{3}{16}u^2+\frac{1}{64}u^3\right)...\right)\\
&{\sim}\left(2+\frac{3}{2}u\right)\left(1+\frac{3}{4}u\right)\\
&{\sim}\left(2+3u\right)\\
\left(\frac{1}{1+\frac{1}{4}u}\right)^{3/2}&=\left(1-\frac{1}{4}u+\frac{1}{16}u^2...\right)^{3/2}\\
&{\sim}\left(1-\frac{1}{4}u\right)^{3/2}\\
&=1^{3/2}-(3/2)\frac{1}{4}u+\frac{1}{2}(3/4)\frac{1}{16}u^2...\\
&{\sim}\left(1-\frac{3}{8}u\right)\\
\left(\frac{1}{1+\frac{1}{4}u}\right)^{5/2}&{\sim}\left(1-\frac{1}{4}u\right)^{5/2}\\
&{\sim}\left(1-\frac{5}{8}u\right).
\end{split}
\end{equation}
\noindent where ``..." in each case denotes the use of a Taylor expansion. Substituting with these approximations, we arrive at
\begin{equation}
\begin{split}
{\zeta}_{\rm \alpha}&{\sim}\frac{2m}{m_{\rm 3}}\left(\frac{R_{\rm 3}}{a_{\rm 2}}\right)^3\left(1+\frac{3}{2}\frac{a_{\rm 1}^2}{a_{\rm 2}^2}\right)\\
{\zeta}_{\rm \beta}&{\sim}\frac{2m}{m_{\rm 3}}\left(\frac{R_{\rm 3}}{a_{\rm 2}}\right)^3\left(1-\frac{3}{4}\frac{a_{\rm 1}^2}{a_{\rm 2}^2}\right).
\end{split}
\label{modzeta2}
\end{equation}
\noindent Since $\frac{5}{2}{\zeta}$ is the magnitude of displacement at the surface of the third body at the points nearest and furthest to the binary system, it can be seen that the third body is still an approximate sphere despite the application of tidal forces.

Thus, combining Eqs. \ref{deltap2} and \ref{modzeta2}, and comparing with Eq. \ref{q2}, we arrive at
\begin{equation}
\begin{split}
{\Delta}E_{\rm P}&=E_{\rm P,ell,{\alpha}}-E_{\rm P,sph,{\beta}}\\
&=(E_{\rm P,ell,{\alpha}}-E_{\rm P,sph})-(E_{\rm P,sph,{\beta}}-E_{\rm P,sph})\\
&{\sim}{\frac{135}{4}}\frac{Gm^{2}R_{\rm 3}^{5}a_{\rm 1}^{2}}{a_{\rm 2}^{8}},
\end{split}
\label{delta_p_homo}
\end{equation}
\noindent which is the difference in self-gravitational potential energy for the third body at equilibrium tide between states $\alpha$ and $\beta$ for a homogeneous body. Note that, interestingly, it is invariant with the mass (or density) of the third body, and is a function of its radius only. This somewhat counterintuitive result is due to our previous assumptions that all tidal distortions are small - a smaller mass would result in a larger geometric distortion, and below a certain mass threshold (when ${\zeta}{\sim}1$), our assumption of small distortion will simply cease to hold.

It should be noted that, in the derivations above, the third body is assumed to be homogeneous. When the mass of the third body is not homogeneously, but only spherically symmetrically, distributed, as is the case for many models of celestial bodies, the equation corresponding to Eq. \ref{deltap1} is
\begin{equation}
\begin{split}
E_{\rm P,ell}-E_{\rm P,sph}&=\int_{0}^{2\pi}\int_{0}^{\pi}\int_{0}^{R_{\rm 3}}|{\Delta}R(\psi,r)|\left({\rho}(r)-{\rho}(r+{\rm d}r)\right)\\
&~~~~~~g(r)({\frac{1}{2}}|{\Delta}R(\psi,r)|)r^{2}\sin{\psi}{\rm d}{\psi}{\rm d}{\xi}\\
\end{split}
\label{nonhomo1}
\end{equation}
\noindent where ${\Delta}R(\psi,r)$ is the vertical displacement of a point mass at ($\psi,r$), and the somewhat elusive ${\rm d}r$ (apparently missing at the end of the expression) is located in the ${\rho}(r+{\rm d}r)$ term. For example, for a body composed of an extremely compact central core and an outer envelope, with the core accounting for 60\% of its mass and the remaining mass being distributed in the envelope according to ${\rho}(r)=kr^{-1.5}$ ($k$ is a constant),
\begin{equation}
\begin{split}
E_{\rm P,ell}-E_{\rm P,sph}&=\int_{0}^{2\pi}\int_{0}^{\pi}\int_{0}^{R_{\rm 3}}|{\Delta}R(\psi,r)|\left({\rho}(r)-{\rho}(r+{\rm d}r)\right)\\
&~~~~~~g(r)({\frac{1}{2}}|{\Delta}R(\psi,r)|)r^{2}\sin{\psi}{\rm d}{\psi}{\rm d}{\xi}\\
&=\int_{0}^{2\pi}\int_{0}^{\pi}\int_{R_{\rm core}}^{R_{\rm 3}}\frac{1}{2}{\Delta}R^{2}(\psi,r)\\
&~~~~\left({\rho}(r)-{\rho}(r+{\rm d}r)\right)g(r)r^{2}\sin{\psi}{\rm d}{\psi}{\rm d}{\xi}\\
&~~~~+\int_{0}^{2\pi}\int_{0}^{\pi}\int_{0}^{R_{\rm core}}\frac{1}{2}{\Delta}R^{2}(\psi,r)\\
&~~~~\left({\rho}(r)-{\rho}(r+{\rm d}r)\right)g(r)r^{2}\sin{\psi}{\rm d}{\psi}{\rm d}{\xi}.
\end{split}
\end{equation}
For a very small core, under the limit of when $R_{\rm core}$ goes to zero, the second term vanishes. To evaluate the first term, one should note that
\begin{equation}
\begin{split}
{\Delta}R(\psi,r)&=\frac{5}{4}r{\zeta}(r)(3{\cos}^2{\psi}-1),\\
{\zeta}(r)&={\zeta}\frac{m_{\rm 3}}{m_{\rm 3}(<r)}\left(\frac{r}{R_3}\right)^3,\\
g(r)&=G\frac{m_{\rm 3}(<r)}{r^2}
\end{split}
\end{equation}
\noindent where ${\zeta}(r)$ is the value of ${\zeta}$ corresponding to a radius of $r$ instead of the surface (i.e. $r=R_3$) of the perturbed body, and $m_{\rm 3}(<r)$ is the total mass included within a sphere of radius $r$ centred at the centre of the perturbed body. It should also be noted that
\begin{equation}
\begin{split}
\left({\rho}(r)-{\rho}(r+{\rm d}r)\right)&=kr^{-1.5}-k(r+{\rm d}r)^{-1.5}\\
&=kr^{-1.5}-kr^{-1.5}\left(1+\frac{{\rm d}r}{r}\right)^{-1.5}\\
&=kr^{-1.5}-kr^{-1.5}\left(1-\frac{3}{2}\frac{{\rm d}r}{r}\right)\\
&=\frac{3}{2}kr^{-2.5}{\rm d}r,
\end{split}
\end{equation}
\noindent and consequently
\begin{equation}
\begin{split}
E_{\rm P,ell}-E_{\rm P,sph}&=\int_{0}^{2\pi}\int_{0}^{\pi}\int_{0}^{R_{\rm 3}}|{\Delta}R(\psi,r)|\left({\rho}(r)-{\rho}(r+{\rm d}r)\right)\\
&~~~~~~g(r)({\frac{1}{2}}|{\Delta}R(\psi,r)|)r^{2}\sin{\psi}{\rm d}{\psi}{\rm d}{\xi}\\
&=\lim_{R_{\rm core} \to 0}\int_{0}^{2\pi}\int_{0}^{\pi}\int_{R_{\rm core}}^{R_{\rm 3}}\frac{1}{2}{\Delta}R^{2}(\psi,r)\\
&~~~~\left({\rho}(r)-{\rho}(r+{\rm d}r)\right)g(r)r^{2}\sin{\psi}{\rm d}{\psi}{\rm d}{\xi}\\
&=\lim_{R_{\rm core} \to 0}\int_{0}^{2\pi}\int_{0}^{\pi}\int_{R_{\rm core}}^{R_{\rm 3}}\frac{1}{2}\frac{25}{16}r^2{\zeta}^2\left(\frac{m_{\rm 3}}{m_{\rm 3}(<r)}\right)^2\\
&~~~~\left(\frac{r}{R_3}\right)^6(3{\cos}^2{\psi}-1)^2\left(\frac{3}{2}kr^{-2.5}{\rm d}r\right)G\frac{m_{\rm 3}(<r)}{r^2}\\
&~~~~r^2\sin{\psi}{\rm d}{\psi}{\rm d}{\xi}\\
&=Gk\frac{75}{64}\lim_{R_{\rm core} \to 0}\int_{0}^{2\pi}\int_{0}^{\pi}\int_{R_{\rm core}}^{R_{\rm 3}}r^{-0.5}{\zeta}^2{m_{\rm 3}^2}\frac{1}{m_{\rm 3}(<r)}\\
&~~~~\left(\frac{r}{R_3}\right)^6(3{\cos}^2{\psi}-1)^2\sin{\psi}{\rm d}r{\rm d}{\psi}{\rm d}{\xi}\\
&=Gk\frac{75}{64}{\times}\left(\int_{0}^{2\pi}{\rm d}{\xi}\right){\times}\left(\int_{0}^{\pi}(3{\cos}^2{\psi}-1)^2\sin{\psi}{\rm d}{\psi}\right)\\
&~~~~{\times}\left(\frac{{\zeta}^2m_{\rm 3}^2}{R_3^6}\right){\times}\left(\lim_{R_{\rm core} \to 0}\int_{R_{\rm core}}^{R_{\rm 3}}\frac{r^{5.5}}{m_{\rm 3}(<r)}{\rm d}r\right)\\
&=Gk\frac{75}{64}{\times}\left(2\pi\right){\times}\left(\frac{8}{5}\right){\times}\left(\frac{{\zeta}^2m_{\rm 3}^2}{R_3^6}\right)\\
&~~~~{\times}\left(\lim_{R_{\rm core} \to 0}\int_{R_{\rm core}}^{R_{\rm 3}}\frac{r^{5.5}}{m_{\rm 3}(<r)}{\rm d}r\right)\\
&=\frac{15{\pi}}{4}k{\frac{Gm_{\rm 3}^{2}{\zeta}^2}{R_{\rm 3}^{6}}}\lim_{R_{\rm core} \to 0}\int_{R_{\rm core}}^{R_{\rm 3}}\frac{r^{5.5}}{m_{\rm 3}(<r)}{\rm d}r.
\end{split}
\label{nonhomo2}
\end{equation}
\noindent To proceed from here, we must ascertain the value of $m_{\rm 3}(<r)$ as a function of $r$, which is
\begin{equation}
\begin{split}
m_{\rm 3}(<r)&=m_{\rm 3,core}+m_{\rm 3,envelope}(<r)\\
&=m_{\rm 3,core}+\int_{0}^{2\pi}\int_{0}^{\pi}\int_{R_{\rm core}}^{r}{\rho}(\tilde{r})\tilde{r}^2\sin{\psi}{\rm d}\tilde{r}{\rm d}{\psi}{\rm d}{\xi},
\end{split}
\end{equation}
\noindent which, under the small $R_{\rm core}$ limit, can be calculated to be
\begin{equation}
\begin{split}
m_{\rm 3}(<r)&=m_{\rm 3,core}+\lim_{R_{\rm core} \to 0}\int_{0}^{2\pi}\int_{0}^{\pi}\int_{R_{\rm core}}^{r}\\
&~~~~~~{\rho}(\tilde{r})\tilde{r}^2\sin{\psi}{\rm d}\tilde{r}{\rm d}{\psi}{\rm d}{\xi}\\
&=\frac{3}{5}m_{\rm 3}+\lim_{R_{\rm core} \to 0}\int_{0}^{2\pi}\int_{0}^{\pi}\int_{R_{\rm core}}^{r}\\
&~~~~~~\left(k\tilde{r}^{-1.5}\right)\tilde{r}^2\sin{\psi}{\rm d}\tilde{r}{\rm d}{\psi}{\rm d}{\xi}\\
&=\frac{3}{5}m_{\rm 3}+4{\pi}k\lim_{R_{\rm core} \to 0}\int_{R_{\rm core}}^{r}\tilde{r}^{0.5}{\rm d}\tilde{r}\\
&=\frac{3}{5}m_{\rm 3}+\frac{8\pi}{3}kr^{1.5}.
\end{split}
\label{nonhomo3}
\end{equation}
\noindent Substituting Eq. \ref{nonhomo3} back into Eq. \ref{nonhomo2}, 
\begin{equation}
E_{\rm P,ell}-E_{\rm P,sph}=\frac{15{\pi}}{4}k{\frac{Gm_{\rm 3}^{2}{\zeta}^2}{R_{\rm 3}^{6}}}\int_{0}^{R_{\rm 3}}\frac{r^{5.5}}{(3/5)m_{\rm 3}+(8\pi/3)kr^{1.5}}{\rm d}r.
\end{equation}
\noindent To calculate this integral, we carry out a numerical integration as follows. Setting $G=1$, $m_{\rm 3}=1$, $\zeta=1$, and $R_3=1$, whereupon $k=\frac{3}{20\pi}$,
\begin{equation}
\begin{split}
E_{\rm P,ell}-E_{\rm P,sph}&=\frac{15{\pi}}{4}k{\frac{Gm_{\rm 3}^{2}{\zeta}^2}{R_{\rm 3}^{6}}}\int_{0}^{R_{\rm 3}}\frac{r^{5.5}}{(3/5)m_{\rm 3}+(8\pi/3)kr^{1.5}}{\rm d}r\\
&=\frac{15{\pi}}{4}\frac{3}{20{\pi}}\int_{0}^{1}\frac{r^{5.5}}{(3/5)+(2/5)r^{1.5}}{\rm d}r\\
&=0.5625{\times}\int_{0}^{1}\frac{x^{5.5}}{0.6+0.4x^{1.5}}{\rm d}x\\
&=0.0940,
\end{split}
\end{equation}
\noindent and hence
\begin{equation}
\begin{split}
E_{\rm P,ell}-E_{\rm P,sph}&=\frac{15{\pi}}{4}k{\frac{Gm_{\rm 3}^{2}{\zeta}^2}{R_{\rm 3}^{6}}}\int_{0}^{R_{\rm 3}}\frac{r^{5.5}}{\frac{3}{5}m_{\rm 3}+\frac{8\pi}{3}kr^{1.5}}{\rm d}r\\
&\sim\frac{1}{20}\left({\frac{15}{8}}{\frac{Gm_{\rm 3}^2{\zeta}^2}{R_{\rm 3}}}\right).
\end{split}
\label{delta_p_layered}
\end{equation}

The density distribution used in the example above is typical of a red giant. Admittedly, real red giant internal density distributions are much more complicated \citep[e.g.][]{1978ApJ...219..183T}, but since the final result is not too sensitive to the index, this is presumably not too bad an approximation. In other words, using a realistic density distribution for $m_{\rm 3}$ will induce a decrease of the self-gravitational potential energy difference, by about an order of magnitude. It can likewise be demonstrated that the final result is not very sensitive to the index of $r$ involved. Again, as with a homogeneous body, the total mass of the receiving body is irrelevant to the result as long as ${\zeta}{\ll}1$, since the $m_{\rm 3}^{2}$ term is cancelled out by the ${\zeta}^2$ term.

It should be noted that these calculations establish only a very generous upper limit of the energy extraction rate of TTs, and should only be regarded as a order-of magnitude estimate of how close a triple system needs to be for TTs to be non-negligible; for an exact calculation, please refer to our Stage 1 simulations in \S 2.

\section{Tidal potential for TTs}
\label{delayed_to_visco}

The gravitational potential of the tertiary is given by expression (\ref{c1}).
Assuming that the shape of the tertiary only departs from a perfect sphere due to the tides raised by $m_1$ and $m_2$, the gravity field coefficients are solely given by the equilibrium tide contribution, i.e., $J_2 = J_2^t$, $C_{22} = C_{22}^t$, and $S_{22} = S_{22}^t$ (Eq.\,(\ref{max2})).
The gravitational potential for coplanar orbits is thus given by
\begin{equation}
\label{gravP}
V({\bm r}) = - \frac{G m_3}{r} + V_1({\bm r}) + V_2({\bm r}) \ ,
\end{equation}
where $V_i ({\bm r})$ is the partial contribution of the mass $m_i$
\begin{equation}
\begin{split}
V_i ({\bm r}) 
&= 
- \frac{G m_3 R_{\rm 3}^2}{2 r^3}\left[k_2 \frac{m_i}{2m_3}\left(\frac{R_{\rm 3}}{r_i}\right)^3\right]\\
&~~~~ - \frac{3 G m_3 R_{\rm 3}^2}{r^3} \left[\frac{k_2}{4}\frac{m_i}{m_3}\left(\frac{R_{\rm 3}}{r_i}\right)^3\cos2{\gamma}_i\right] \cos {2 \gamma}\\
&~~~~ + \frac{3 G m_3 R_{\rm 3}^2}{r^3} \left[-\frac{k_2}{4}\frac{m_i}{m_3}\left(\frac{R_{\rm 3}}{r_i}\right)^3\sin2{\gamma}_i\right] \sin {2 \gamma}\\
&= - \frac{G m_3}{r}\left[k_2 \, \zeta_i \left(\frac{R_{\rm 3}}{r}\right)^2 P_{\rm 2}(\cos({\gamma}_i-\gamma))  \right] , 
\end{split}
\label{appB1} 
\end{equation}
\noindent with
\begin{equation}
\label{rphii}
\zeta_i =\frac{m_i}{m_3}\left(\frac{R_{\rm 3}}{r_i}\right)^3 \ . 
\end{equation}
For ${\bm r}$ in the orbital plane, we additionally have
\begin{equation}
\label{cospsii}
\cos ({\psi_i}-\psi) = \cos ({\gamma}_i-\gamma)  \ , 
\end{equation}
and we can rewrite the gravitational potential (\ref{gravP}) as
\begin{equation}
V({\bm r})=-\frac{Gm_{\rm 3}}{r} \left[1+k_{\rm 2} \, \zeta \left(\frac{R_{\rm 3}}{r}\right)^{2}P_{\rm 2}(\cos{\psi}) \right].
\label{appBphi}
\end{equation}
\noindent For a single perturber, for instance $m_1$, we have $\zeta = \zeta_1$ and $\psi_1 = 0$, so expression (\ref{appBphi}) gives the usual tidal potential (since $m_2=0 \Rightarrow \zeta_2=0$). For two perturbers, $\zeta$ depends on the relative position of these perturbers with respect to the tertiary, and so we have
\begin{equation}
\zeta \, P_{\rm 2}(\cos{\psi}) = \zeta_1 P_{\rm 2}(\cos({\psi_1}-\psi)) + \zeta_2 P_{\rm 2}(\cos({\psi_2}-\psi)) \ .
\end{equation}

To simplify things, we can approximate $m_{\rm 3}$ as an ellipsoid with its singular bulge constantly pointed towards the inner binary COM. In other words, we assume that the respective tidal bulges raised by $m_{\rm 1}$ and $m_{\rm 2}$ can be approximated to coalesce to from a single set of bulges, equal to that raised by a point mass at the COM of the inner binary. In this scenario, we only need to find the value of $\zeta$ for any value of $\psi$ in order to find the $\zeta$ that characterises the deformation of the entire $m_{\rm 3}$, regardless of $\zeta$. Therefore, setting $\psi=0$, and noting that $m=m_{\rm 1}=m_{\rm 2}$, we have
\begin{equation}
\begin{split}
\zeta  &= \zeta_1 P_{\rm 2}(\cos{\psi_1}) + \zeta_2 P_{\rm 2}(\cos{\psi_2})\\
&=\frac{m}{m_3}\left(\frac{R_{\rm 3}}{r_1}\right)^3P_{\rm 2}(\cos{\psi_1}) + \frac{m}{m_3}\left(\frac{R_{\rm 3}}{r_2}\right)^3P_{\rm 2}(\cos{\psi_2})\\
&=\left[\frac{P_{\rm 2}(\cos{\psi_1})}{(r_{\rm 1}/a_{\rm 2})^3}+\frac{P_{\rm 2}(\cos{\psi_2})}{(r_{\rm 2}/a_{\rm 2})^3}\right]\frac{m}{m_{\rm 3}}\left(\frac{R_{\rm 3}}{a_{\rm 2}}\right)^3.
\end{split}
\end{equation}
\noindent which is exactly Eq. \ref{zetaphi}.


\bsp	
\label{lastpage}
\end{document}